# Laboratory simulation of field aligned currents in experiment on laser-produced plasma interacting with magnetic dipole

**I F Shaikhislamov, V M Antonov, Yu P Zakharov, E L Boyarintsev, A V Melekhov, V G Posukh and A G Ponomarenko**

Dep. of Laser Plasma, Institute of Laser Physics SB RAS, pr.Lavrentyeva 13/3, 630090, Novosibirsk

E-mail: ildars@ngs.ru

**Abstract**
In experiment on magnetic dipole interacting with laser-produced plasma a generation of intense field aligned current (FAC) system was observed for the first time in a laboratory. Detailed measurements of total value and local current density, of magnetic field at the poles and in the equatorial magnetopause, and particular features of electron motion in the current channels revealed its similarity to the Region-1 current system in the Earth magnetosphere. Such currents were found to exist only if they can closer via conductive cover of the dipole. Comparison of conductive and dielectric cases revealed specific magnetic features produced by FAC and their connection with electric potential generated in the equatorial part of magnetopause. To interpret data we consider a model of electric potential generation in the boundary layer which agrees with experiment and with measurements of the Earth' transpolar potential in the absence of interplanetary magnetic field as well. Results could be of importance for investigation of the Mercury as a magnetic disturbance due to FAC could be especially large because of small size of the Hermean magnetosphere.

## 1. Introduction

The field aligned current is a key element in the Earth magnetosphere-ionosphere coupling. The magnetospheric currents can flow into the ionosphere through the Birkeland currents, which were described for the first time by (Iijima and Potemra 1976), and closer through the ionosphere transverse to magnetic field due to finite Pedersen conductance. Region-1 current at the dayside maps in the Plasma Sheet Boundary Layer and is a direct result of the Solar Wind interaction with the Earth magnetic field. In the dawn sector current is carried by upward accelerated bursts of electrons extracted from the ionosphere, while in the dusk sector by electrons precipitating from the PSBL (one of the recent measurements by Cluster is reported in Marchaudon *et al* 2006). There is also a Region-2 current of reverse polarity observed at lower latitudes that maps into the Central Plasma Sheet inside the magnetosphere. Multiple observations of transpolar potential which relates to intensity of Region-1 current revealed a noticeable dependence on direction of Interplanetary Magnetic Field. For the southward direction of typical IMF the potential is around three times larger than for the northward (for example, Shepherd 2007).

In theoretical analyses inhomogeneity and curvature of magnetic field, plasma pressure gradient and viscosity are considered as drivers of FAC (Hasegawa and Sato 1979, Vasyliunas 1984, Itonaga *et al* 2000). Dependence of magnetosphere-ionosphere coupling on IMF is attributed to reconnection at the dayside magnetopause. Empirical model of (Hill 1984) developed further by (Siscoe *et al* 2002a) explains saturation of transpolar potential by a feedback influence of FAC on plasma motion and magnetic field near the magnetopause. For this mechanism a maximum value of total field aligned current depends on wind pressure and magnetic moment as $\sim (p \cdot \mu)^{1/3}$. During the last two decades a number of investigations based on MHD numerical simulation were performed, such as on influence of ionosphere conductance on the properties of Solar wind interaction with magnetosphere (Fedder and Lion 1987), impact of energetic CME plasma on the Earth magnetosphere and ionosphere (Ridley *et al* 2006) and others. In most MHD simulations that take into account ionospheric conductance a generation of Region-1 current was observed as well as their significant increase following dynamic pressure jumps in the Solar Wind.

Discovery of intrinsic magnetic moment of the Mercury by Mariner in 1974 opened up new possibilities in the physics of planetary magnetospheres. From spacecraft measurements a presence of FAC on nightside was also deduced (Slavin *et al* 1997), and the inferred current turned to be unexpectedly large and comparable to that of the Earth, ~ 1 MA. Ground based observations revealed a sporadic Sodium luminosity on planet surface concentrated at high latitudes (Potter and Morgan 1990) which is attributed to magnetospheric processes. In view of soon to be operating MESSENGER and planned BepiColombo missions, the influence of FACs on structure and dynamics of magnetosphere was formulated as one of the main problems in Mercury investigations (Baumjohann *et al* 2006). Due to relatively small size and large gyroradius of exsosphere ions laboratory simulation of Hermean magnetosphere looks promising.

So far in terrella laboratory experiments FAC generation was studied only in the works of (Rahman *et al* 1991). Magnetic features characteristic for the Earth regions (1, 2 and NBZ) were found over dipole poles. Intensity and spatial distribution of magnetic perturbation depended on the direction of "IFM". However, presence of FAC was inferred only from magnetic measurements. Magnetic perturbation $\Delta B_x$ in the sunward direction (opposite to plasma flow) was several times smaller than other components. Dipole didn't have conductive cover so current closer wasn't obvious. The authors of the present work also carried out terrella experiments with approximately the same parameters (Ponomarenko *et al* 2002). At a certain stage of interaction, well after the quasi-stationary magnetopause was formed, a restructuring of magnetic field over poles was observed. The vector of magnetic perturbation from mainly vertical direction rotated toward plasma flow, which is to be expected if Region-1 current develops. Absolute and relative value of $\Delta B_x$ component was one order of magnitude larger than in the cited above experiment.

Thus, field aligned current as a specific phenomenon is observed in markedly different conditions of space and laboratory plasmas and in numerical simulations as well. Despite of overwhelming amount of data and knowledge development the main questions on prevailing mechanism of FAC generation and in what region of magnetopause remain largely unanswered. Laboratory simulation can offer a novel point of view and qualitatively new data inaccessible by other means. The paper describes results of such complex laboratory investigation. In the experiment a dipole with large moment and correspondingly large size was used, while the wind was modeled by energetic flow of laser-produced plasma. In the present work there is no magnetic field frozen into plasma. Investigation concerns only the dayside magnetosphere as no tail forms during a flow time-span of laser-produced plasma. Despite of these simplifications and relatively brief interaction time, formation of well defined magnetopause and cusps, and generation of intense FAC were observed, while larger scale of magnetosphere made it possible to find out and study details which are too small in terrella experiments.

Fundamental aspects of magnetic dipole interaction with explosive plasma have been analyzed in (Nikitin and Ponomarenko 1995). In the MHD frame there is a single energetic parameter $\chi = 3W_o R_o^3 / \mu^2$ that binds magnetic moment value $\mu$ with energy $W_o$ of explosion taking place at a distance $R_o$ from the dipole center. At $\chi \gg 1$ the overflowing kind of interaction realizes, while at $\chi \leq 1$ plasma is captured as a whole. Experiments in the regime $\chi \gg 1$ model extreme compression of the Earth magnetosphere by plasma flow of powerful CME (Ponomarenko *et al* 2007, 2008, Zakharov *et al* 2008). Processes in polar regions were investigated previously by the authors in the regime $\chi \leq 1$ which models releases inside magnetosphere (Antonov *et al* 2001). Particle and energy fluxes at poles, transpolar potential and field aligned current were measured for the first time.

In the present work experiment is realized in the overflowing regime with energetic parameter $\chi \sim 100$. Total FAC value and a local current density were measured. Comparison of conductive and dielectric cover shields of the dipole allowed for the first time to clearly distinguish the FAC input into magnetospheric field not only over the poles but at the equator as well. Current closer in the dipole cover gives rise to a specific luminosity at footprints. Its spectral analysis gave information on current carriers and electron motion which in general outlook are similar to observations in the ionosphere. Presented data on FAC dependence on the dipole moment reveal the Chapman-Ferraro scaling. A link between FAC and electric potential in the equatorial magnetopause was found out. The paper consists of two sections on experimental results and their discussion including model for data interpretation, followed by conclusions.

## 2. Experimental set up and results

Experiment has been carried out at KI-1 space simulation Facility, which includes chamber 5 m in length and 1.2 m in diameter with operating base pressure $10^{-6}$ Torr. As shown in figure 1, two $CO_2$ beams of 70 ns duration and 150 J of energy each were focused and overlapped into a spot 1 cm in diameter on surface of a solid target. The target was made of perlon ($C_6H_{11}ON$) in form of a semi-sphere with a radius of 3 cm. Laser-produced plasma consisted mostly of $H^+$ and $C^{4+}$ ions approximately in equal parts and expanded inertially in a cone ≈1 radian with an average velocity $V_o \approx 1.5 \cdot 10^7$ cm/sec. A total kinetic energy and a number of ions in the flow was $W_o \approx 40$ J and $N_o \approx 5 \cdot 10^{17}$ respectively. Due to specific pulse and tail generation mode of laser oscillator there was, besides the main plasma flow, a secondary plasma twice as slow and order of magnitude less energetic. Its influence on the magnetospheric cavity was negligible but was more pronounced in the polar regions. At the axis of plasma expansion at a distance of $R_o=70$ cm magnetic dipole was placed. A maximum value of magnetic moment was up to $\mu=1.2 \cdot 10^7$ Gs·cm$^3$; a fall off time $\sim 10^{-3}$ sec while a typical time of interaction duration $\sim 10^{-5}$ sec. The dipole has a stainless cover shield in form of a cylinder 20 cm in diameter and 16 cm in height.

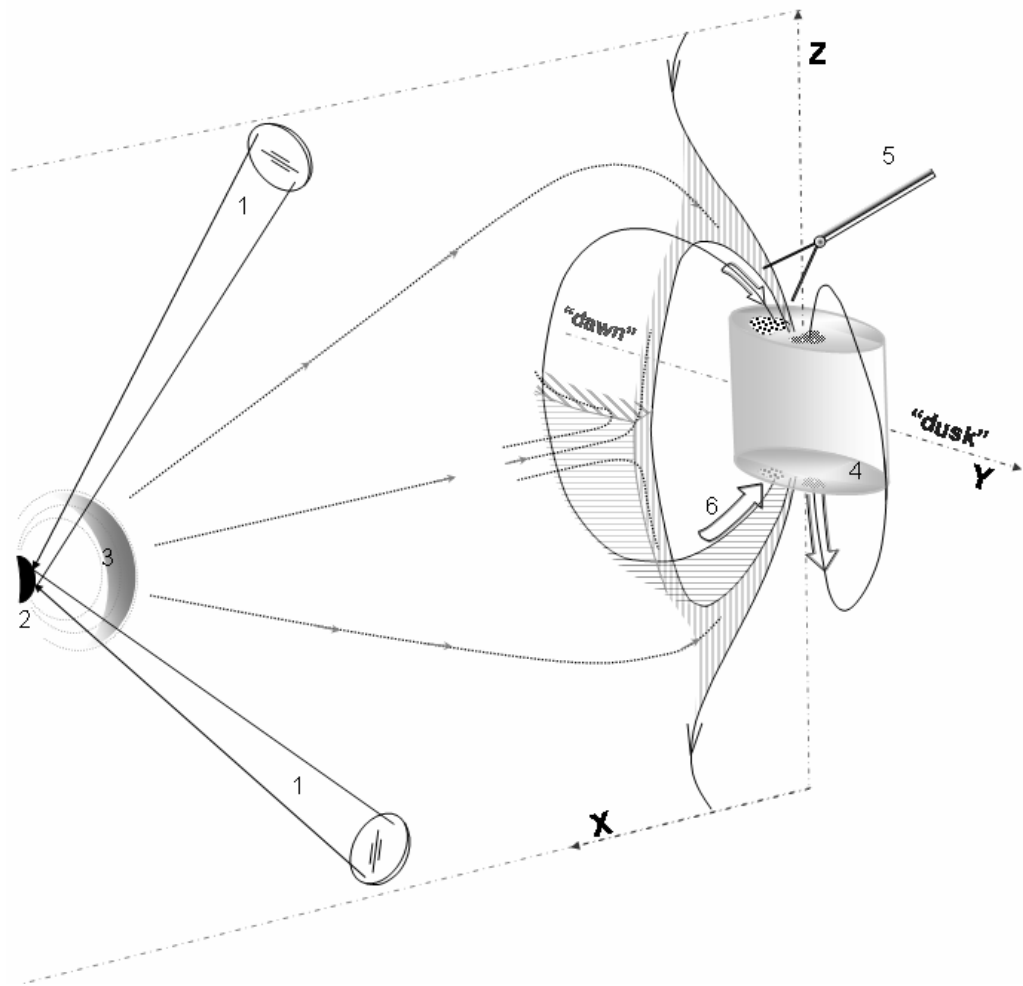

**Figure 1.** Experimental set up. 1 – laser beams; 2 – target; 4 – magnetic dipole; 5 – probes. Also schematically are shown laser-produced plasma (3) and field aligned currents (6, large arrows). Thin lines mark magnetic field lines and plasma streamlines.

In the present experiment magnetic moment was always oriented perpendicular to the interaction axis. We use GSM coordinate frame throughout the paper, as shown in figure1. Diagnostics consisted of miniature electric and three-component magnetic probes, gated imaging with an exposition of $3 \cdot 10^{-8}$ sec, optical spectroscopy of plasma radiation, Rogovski coil of 1.7 cm radius to measure current density in plasma. Some of dimensionless parameters of the experiment are listed in table.

**Table 1.** Dimensionless parameters of experiment.

| Parameter | Lab. | Earth | Mercury |
|---|---|---|---|
| Magnetopause size $R_m$ to the dipole radius | 2÷3 | ~10 | ~1.5 |
| Degree of ion magnetization $R_L/R_m$ | 0.3 | ~$10^{-3}$ | ~$10^{-2}$ (for Sodium ~0.2) |
| Knudsen number $\lambda_i/R_m$ | ~5 | >>1 | >>1 |
| Reynolds number $4\pi\sigma R_m V/c^2$ | ~5 | >>1 | >>1 |
| Hall parameter $4\pi en V R_m / cB$ | ~5 | >>1 | >>1 |

At a time of about t=2 μsec after laser irradiation of the target plasma reaches a deceleration region and at a distance of $R_m$=20÷30 cm, depending on conditions, a well defined magnetopause is formed. It divided the external region of zero or weak magnetic field from the inner magnetosphere of strong field and remained in place for about 3 μs, after which moved away from the dipole. On a typical meridian snapshot (figure 2) one can clearly see the form of magnetopause and the cusps. Another distinct feature was a bright structured luminosity at each of the dipole poles. It was observed in a number of other experiments at KI-1 (Antonov *et al* 2001). On a typical snapshot of the polar region, shown in figure 3, a pair of spots located at latitudes 45÷65° at the opposite sides relative to the direction of plasma flow is noticeable.

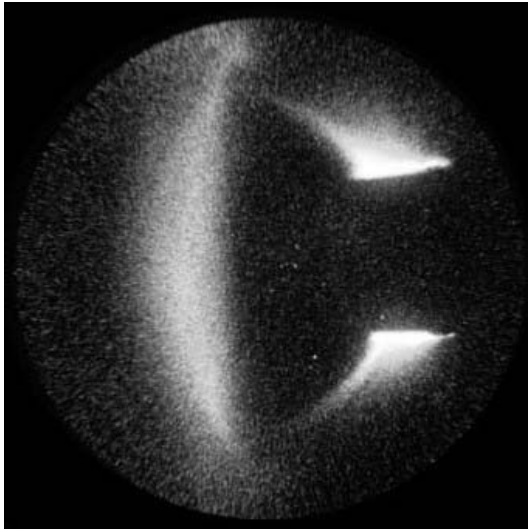
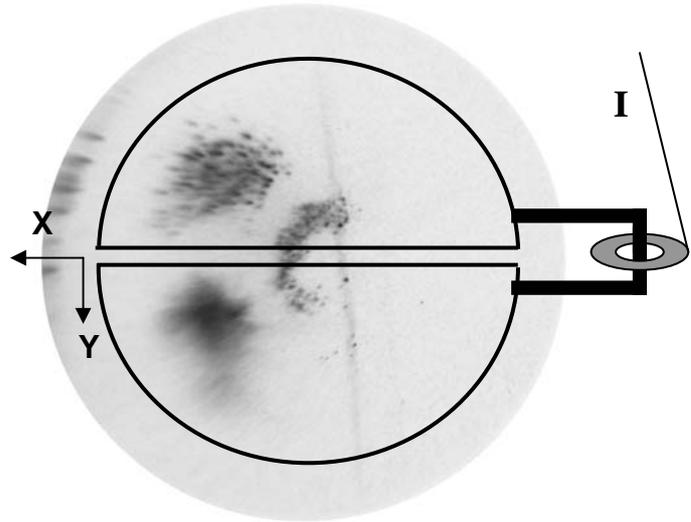

**Figure 2.** Typical meridian snapshot of laser-produced plasma interacting with dipole.

**Figure 3.** Snapshot of the dipole polar region at a time of t=3 μsec. Also shown is a scheme of current measurement.

Further on the geophysics terms of dawn and dusk sides will be used which in a laboratory correspond to the different signs of a vector product $\vec{r} \cdot [\vec{\mu} \times \vec{V}_o]$. The dawn spot (Y<0) in comparison with the dusk spot (Y>0), consists of small intense blots with attached thin filaments spread along the field lines and has a different colour. First appearance of the spots coincides with plasma arriving at the dipole surface and is followed by increase in size and shifting in latitude and longitude. Scanning of plasma snapshots (figure 2) along the X-axis and comparison with magnetic measurements showed that luminosity well coincides spatially with the magnetosheath. Sharp boundaries correspond to a minimum and a maximum of magnetic perturbation, while a maximum of luminosity – to a maximum density of Chapman-Ferraro current. In figure 4 profiles of plasma density and magnetic perturbation measured by probes along the interaction axis are shown.

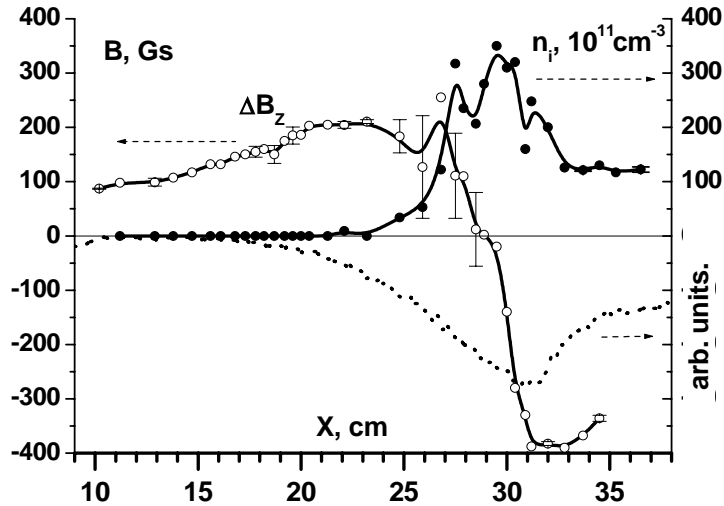

**Figure 4.** Profiles of magnetic field perturbation (o), plasma density (●) and plasma luminosity (dotted curve) along the interaction axis obtained at a time of t≈2.25 μsec. Magnetic moment of the dipole μ=$10^7$ Gs·cm$^3$.

One can see density jump, magnetosheath, field compression in the inner region and total field expulsion outside the boundary layer. A width of the boundary layer δ=3÷5 cm is close to the ion inertia length $c/\omega_{pi}$ ≈4÷6 cm. After forming of magnetopause a reflected ion flow was detected upstream of the layer with flux intensity approximately equal to the initial one. A scaling of magnetopause size $R_m$ on the dipole moment measured by characteristic points of plasma luminosity and magnetic field profiles is presented in figure 5.

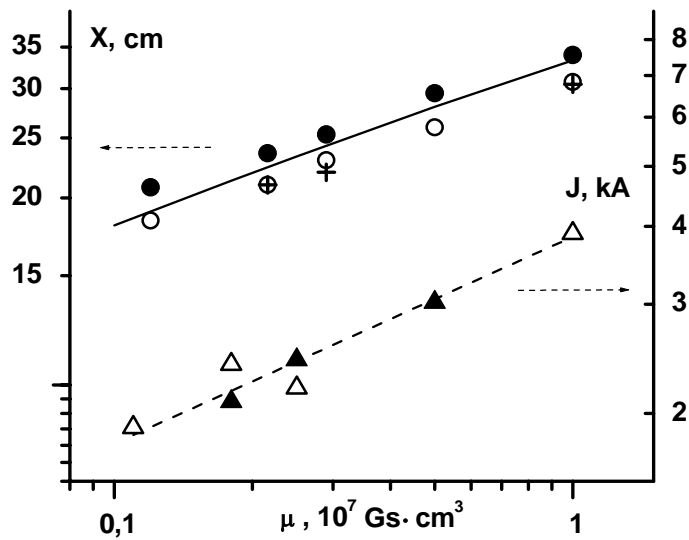

**Figure 5.** Dependence of magnetopause position on the dipole moment plotted by maximum of plasma luminosity (o), maximum gradient of luminosity (●) and maximum gradient of ΔB$_z$ (+). Solid line – analytical calculation.
Right abscissa – value of total field aligned current measured between plates (▲) and by magnetic probes (Δ). Dashed line - dependence ~μ$^{1/3}$.

The position of magnetopause is determined by a pressure balance, $m_i n_i V^2 = k\mu^2/8\pi R_m^6$, where numerical coefficient $k$ depends on specifics of ion reflection and geometry. Dynamic pressure of expanding laser-produced plasma can be expressed through initial kinetic energy of explosion, $m_i n_i V^2 = 6\cdot(\partial W_o/\partial\Omega)/(R_o - R_m)^3$, where $R_o$ is a distance from the dipole center to the point of plasma origin. Dependence $R_m(\mu)$ calculated taking into account parameter $R_m/R_o$ and measured value $\partial W_o/\partial\Omega$ of energy per solid angle is shown in figure 5 by solid line.

Drawing magnetic field lines using analytical model for Chapman-Ferraro currents revealed that the polar spots (figure 3) map in the magnetosheath and density jump and adjacent inner magnetosphere shown in figure 4. Besides of the spots there was observed also a thin high-latitude oval encircling the pole axis. This structure is probably related to a high-latitude boundary layer on the other side of cusp where Chapman-Ferraro current flows in opposite direction relative to the equatorial part. The cusp itself maps between the oval and the spots at latitudes $60 \div 70^\circ$.

The current that flows in the dipole cover through a noon-meridian cut was measured by a device shown in figure 3. Thin aluminum or copper plates electrically detached from the dipole cover and separated from each other along the interaction axis were placed over the pole. A shortcutting shunt measured total current flowing between plates. Local current density in plasma was measured by Rogovski coil. The coil was oriented parallel to polar surface and could move over the dawn spot at a fixed height of 5 cm. Figure 6 shows corresponding oscilloscope signals. Local current was inhomogeneous in space with 100% modulation. Averaging yielded that it was concentrated mostly in a region of the luminous spot. In figure 6 the presented measurement was obtained at the dawn spot center where signal was maximal. At coil crossing of dividing line between the spots a signal changed polarity. From presented measurements it follows that at the dawn side current flows into the dipole cover, then flows inside the cover in the dawn-dusk direction and finally flows away from the cover at the dusk side. Estimation of total current by a sum of local measurements made over the pole gives $\approx 1.3$ kA which is around one and a half times less than the current between the plates. This discrepancy is probably attributed to instrumental error and incompleteness of Rogovski coil measurements.

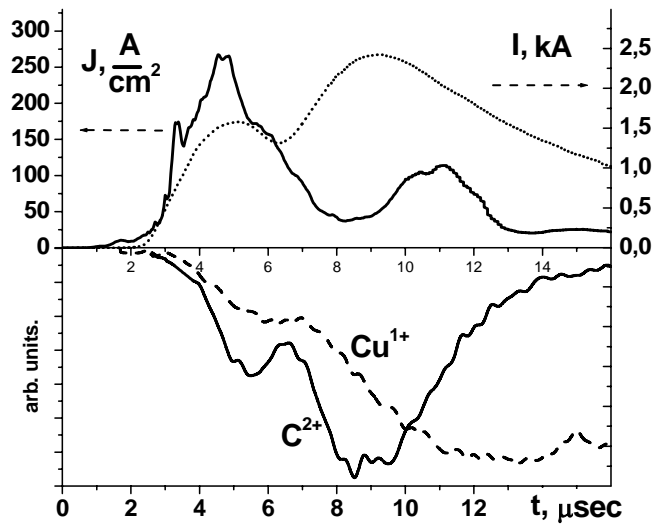

**Figure 6.** Upper panel – oscilloscope signals of local current density (solid) and total current between plates (dotted).
Lower panel – luminosity of line 464.7 nm of $C^{2+}$ ion in the dusk spot (solid) and line 505 nm of $Cu^{1+}$ ion in the dawn spot (dashed). $\mu = 0.25 \cdot 10^7$ Gs·cm$^3$.

In figure 6 there are also shown signals of spectral luminosity of plasma ions (line $\lambda = 464.7$ nm, $C^{2+}$) and plate material ions (505 nm, $Cu^{1+}$) observed at the pole surface. Time behavior of all values is similar and reflects arriving at poles of two plasma flows at $\approx 5$ and 9 μsec. It should be noted that the luminosity of plate material ions after being exited was distinctly more lasting, up to 40 μsec. Detailed spatially and temporally resolved spectral measurements in optical range revealed the following.

Radiation of plasma ions, which consisted of a few strong lines, could be seen in a wide oval embracing spots and a space between them. However, intensity of lines was order of magnitude higher in the dusk spot than elsewhere. Radiation of atoms and singly ionized ions of plate material was concentrated in small point-like areas which form the dawn spot. The same picture was observed on stainless cover of the dipole proper. In this case lines of Fe could be seen. The process that induces luminosity of surface material is an explosive emission of electrons and their acceleration into plasma. A presence of positive potential in plasma over the dawn spot was found out in previous experiment (Antonov *et al* 2001). Over the dusk spot electrons move from plasma into metal and knock out cold secondary electrons which in turn recombine with plasma ions and induce their luminosity.

Magnetic perturbation at poles produced by FAC was measured by probe that could cross the pole region in meridian plane at an angle of $45^{\circ}$ to X and Z axis with a minimal distance to the dipole center 14 cm. The vector of magnetic field perturbation was found to be directed mainly along X axis, while z-component was several times smaller. The probe detected also $\Delta B_y$ component because its path passed closer to the dawn than the dusk spot. At crossing the $\Delta B_y$ component changed sign as shown in figure 7. From $\Delta B_y$ and $\Delta B_x$ profiles a total current $I \approx 1.9$ kA and a current channel diameter $\approx 3$ cm could be calculated independently of other measurements. In figure5, besides of magnetosphere size, a value of total FAC derived from direct measurements of current between plates and from $\Delta B_y$ profiles is shown. Dependence on the dipole moment is close to $\sim \mu^{1/3}$ scaling.

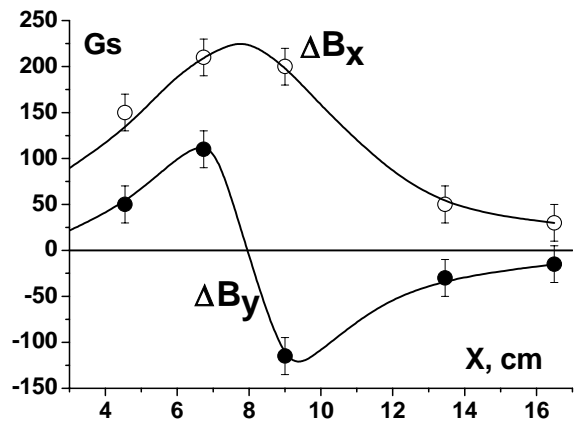

**Figure 7.** Profiles of magnetic field perturbation measured by probe crossing over the dawn spot at the north pole. $\mu = 0.18 \cdot 10^7$ Gs·cm$^3$.

To clarify the input of FAC into the inner magnetospheric field we compare results obtained for conductive and dielectric covers of the dipole while other parameters were kept the same. Dielectric film covering whole dipole surface effectively suppressed FAC. Namely, luminosity at the poles, in lines as well as integral over visible range, was tens times weaker and didn't have a double spot structure. Local current density didn't exceed signal noise level, and $\Delta B_y$ component was small and didn't show sign reversal while probes crossed over the pole, unlike figure 7. In figure 8-a time behavior of magnetic field components measured over the north pole at a point (X=8.5, Z=9.5, Y=0 cm) is presented. For the case of conductive cover curves are marked by a capital letter *B*, while for the dielectric case – by a small *b*. Close to a noon-meridian plane $\Delta B_y$ component was small and isn't shown. When FAC flow a field perturbation dynamic is similar to that of currents (figure 6). Two maxima at $\approx 3.5$ and 8.5 μsec correspond to arrival of plasma at the poles. When FAC is absent the field perturbation over the pole is significantly smaller, while dynamic relates to plasma arrival at the equatorial magnetopause, rather than poles. Note also that z-component has an opposite sign as well. Time behavior at the equator inside of magnetosphere at a point (X=14, Z=0, Y=0 cm) is shown in figure 8-b. One can see that when FAC is absent a change of the main component $\Delta b_z$ is always positive which is to be expected if there is only Chapman-Ferraro current. FAC decreases field compression by about 30% and at later time perturbation of the main component at the equator becomes even negative.

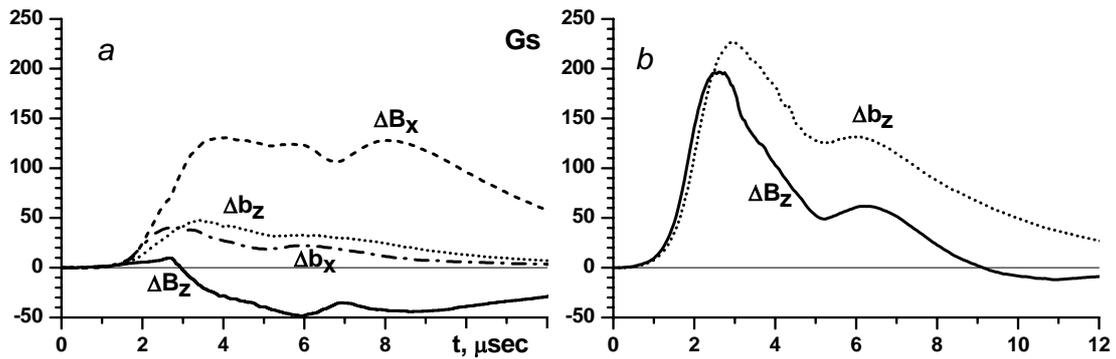

**Figure 8.** Oscilloscope signals of magnetic field perturbation over the north pole (*a*) and at the equator (*b*) for the cases of conductive (solid for $\Delta B_z$; dashed for $\Delta B_x$) and dielectric (dotted for $\Delta b_z$; dash-dot for $\Delta b_x$) dipole cover.

The difference between two cases is clearly demonstrated in figure 9 by equatorial profiles of magnetopause measured at the same time. While position of the magnetosheath isn't affected by FAC, magnetic field perturbation inside of magnetosphere noticeably changes.

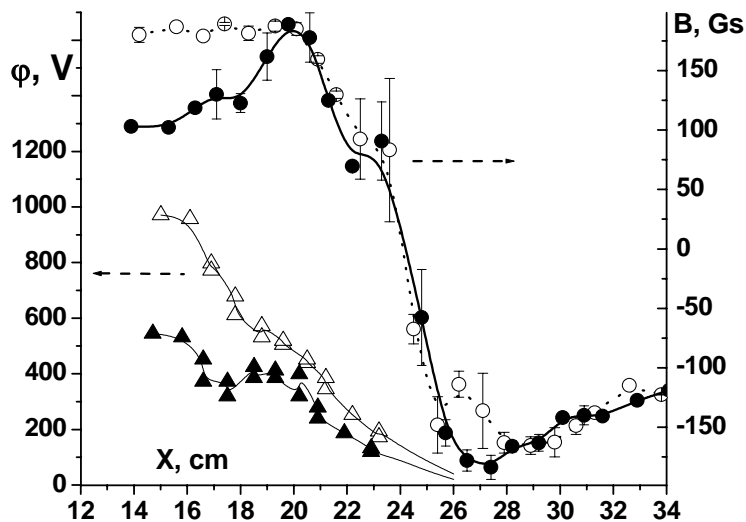

**Figure 9.** Right abscissa – profiles of magnetic field perturbation along X-axis measured at a time of t=4 μsec in the case of conductive (●) and dielectric (○) dipole cover. Left abscissa – profiles of electric potential in the conductive (▲) and dielectric (Δ) case. $\mu=0.25 \cdot 10^7$ Gs·cm$^3$.

Measurements by Langmuir probes revealed that across the magnetosheath there is electric potential in plasma as shown in figure 9. Its value was systematically larger at the dawn than at the dusk side. This is demonstrated in figure 10 by signals obtained at the equatorial magnetopause (Z=0, X=19 cm) at three different locations in respect to Y coordinate. Maximum value of the dawn-dusk potential drop was measured to be 0.5÷1 kV. Note that it reaches maximum at a time of plasma flux maximum in the equatorial region t≈2.5 μsec and rapidly falls off after plasma arrives at poles and FAC develops at around t≈5 μsec. A second maximum in figure 10 is generated by the second plasma flow.

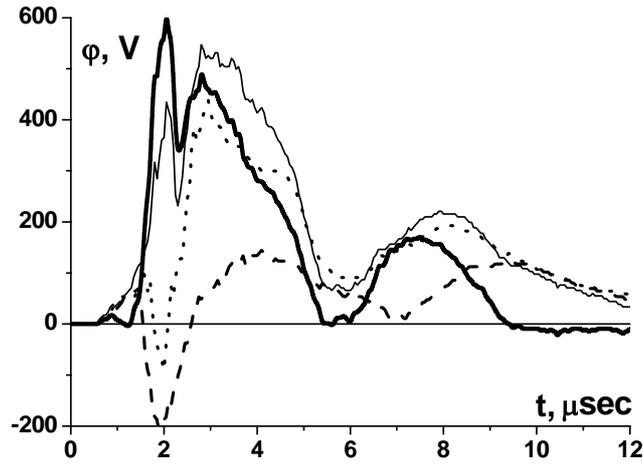

**Figure 10.** Oscilloscope signals of electric potential measured at three points of equatorial magnetopause Y=-6.2 cm (thin solid), Y=-0.8 cm (dash) and Y=4.5 cm (dotted). Bold curve shows a difference between two extreme points.

## 3. Discussion

As a whole obtained data give strong evidence that in laboratory experiments on dipole interaction with laser-produced plasma an intense field aligned currents exist. They flow in a sense of Region-1 current. In laboratory current closer and FAC existence is provided by highly conductive dipole cover and explosive emission of electrons from metal into plasma at dawn sector. It appears that current in polar regions is linked with magnetosheath at equator and with inner magnetosphere adjacent to magnetopause. Namely in those regions a marked differences in electric potential were observed. For nonconductive "ionosphere" it was measured to be up to 1 kV, while for conductive one potential decreased by several times after FAC development. This indicates that FAC might actually originate in low-latitude rather high-latitude boundary layer. A decrease of $\Delta B_z$ component of field perturbation in the inner magnetosphere section (X=17÷20 cm, figure 9) is explained mostly by field line curvature, and corresponding positive and negative $\Delta B_x$ components were observed above and down of the equator. Thus, FAC loop, if it originates in equatorial boundary layer, must be more complex than a simple quasi-circle.

Observed dependence of total FAC value on dipole moment indicates on its MHD origin. The total current measured and accounted for both poles ~4 kA is comparable to total Chapman-Ferraro current in equatorial magnetopause ~8 kA. We note that a maximum estimate of FAC for the Earth (Siscoe et al 2002b) and Mercury yields the same relation. Estimating Chapman-Ferraro current as $I_{CF} \sim B_m R_m$, magnetic field of FAC current as $B_{FAC} \sim I_{FAC}/R_o$, equating $I_{CF} \approx 2 \cdot I_{FAC}$ and expressing field at magnetopause through dipole field at poles $B_m \approx 0.5 \cdot B_o (R_o/R_m)^3$ we obtain estimation of field perturbation at poles expressed in terms of magnetopause and planet sizes only $B_{FAC} \cong 0.25 \cdot B_o (R_o/R_m)^2$. One can see that in Hermean magnetosphere FAC input could be especially large, up to 10% of the main dipole field. The given formula yields a value of 400 Gs for laboratory experiment which is to be compared with measured 200÷400 Gs, 100 nT for the Earth while observed values are 100÷500 nT, and ~50 nT for the Mercury. Thus, a precise calculation of Hermean magnetic moment from the available and future spacecraft data needs to take into account a possible FAC input. We note that so far a discrepancy on the level of a few tens of nT exists between equatorial and high latitude spacecraft measurements (Anderson et al 2008).

From many existing models most simply results presented in figure 9 and 10 explains the model of electric potential generation in a boundary layer, and there are a number of woks on this topic (for example, Denisenko et al 1992 and references herein). Let's consider a generalized Ohm' law that follows from momentum equation for electrons with electron mass ignored:

$$\vec{E} + \frac{\vec{V} \times \vec{B}}{c} - \frac{\vec{J} \times \vec{B}}{nec} + \frac{\vec{\nabla} P_e}{ne} - \frac{\vec{J}}{\sigma} = 0 \qquad (1)$$

Electric field has an induction part and a potential part $\vec{E} = -\vec{\nabla}\varphi$. We assume that when plasma overflows dipole a thin boundary layer forms which separates external flow from inner magnetosphere due to Chapman-Ferraro current and which maps along open field lines via cusps in polar regions. Inside the layer a product $\vec{V} \times \vec{B}$ is not zero, and its width is determined by viscosity or anomalous diffusion. As all values change much faster across the layer than along it, roughly along X-axis, a potential part of (1) $\nabla^2 \varphi = \vec{\nabla} \cdot \vec{E}$ could be written as:

$$\frac{\partial^2 \varphi}{\partial x^2} \approx \frac{\partial}{\partial x}\left(\frac{1}{c} V_y B_z - \frac{J_y B_z}{nec} + \frac{1}{ne}\frac{\partial P_e}{\partial x}\right) \qquad (2)$$

One can see that the convection term $V_y B_z$ has dawn-dusk asymmetry that should be compensated by asymmetry of potential. Ignoring field line curvature, approximate solution follows as:

$$\varphi \approx -\int_x^\infty \left[\frac{1}{ne}\frac{\partial}{\partial x}\left(B^2/8\pi + P_e\right) + \frac{1}{c}V_y B_z\right] \cdot dx \qquad (3)$$

Thus, due to a pressure balance at the magnetopause $B^2/8\pi \approx m_i n_i V^2$, potential across the boundary layer or magnetosheath has a typical value of ion kinetic energy. It is larger on the dawn side where $V_y < 0$. Assuming that the flanking velocity $V_y$ doesn't exceed downstream velocity, maximum dawn-dusk potential drop could be estimated as:

$$\Delta\varphi \sim 2 V_o B_z \delta/c \qquad (4)$$

For experimental conditions magnetosheath width $\delta$ is comparable to ion gyroradius, so $\Delta\varphi$ estimates as ion kinetic energy, or $\approx 1$ kV. Width of the Earth magnetosheath varies in a wide range 100÷1300 km with an average value 400÷800 km (by Cluster data, Paschmann *et al* 2005). This is by order of magnitude larger than gyroradius and respective $\Delta\varphi$ estimates as ~10 kV. According to statistical observations, transpolar potential in absence of IFM is proportional to square velocity of SW and for $V_o$=300 km/sec is about $\approx 9$ kV (Boyle *et al* 1997). We note that in (Denisenko *et al* 1992) a value of $\Delta\varphi \approx 8$ kV was derived by means of numerical calculation.

Electric potential is translated along field lines from the low latitudes into the dipole poles. If the dipole surface is conductive, then current should exist. Its value would depend on the sum of integrated cross-field ionosphere conductance and conductivity of plasma column above ionosphere. If the total conductance is large enough, for example in the experiment it was $\geq 1$ Ohm$^{-1}$, a problem of current saturation arises. However, actual current generation should be calculated by induction part of (1), $\partial \vec{J}/\partial t \sim -\vec{\nabla} \times \vec{\nabla} \times \vec{E} = \nabla^2 \vec{E} - \vec{\nabla}(\vec{\nabla} \cdot \vec{E})$. As $\nabla^2$ is a scalar Laplas operator, current along field lines is generated by the same terms at the right side of (2) that give rise to electric potential:

$$\frac{\partial J_z}{\partial t} \approx \frac{c^2}{4\pi}\frac{\partial}{\partial z}\frac{\partial}{\partial x}\left(\frac{1}{c}V_y B_z - \frac{J_y B_z}{nec} + \frac{1}{ne}\frac{\partial P_e}{\partial x}\right) \qquad (5)$$

These simple arguments show how FAC and potential drop are linked through the asymmetric convection term $V_y B_z$. Extensive analysis of parallel current generation could be found in (Itonaga *et al* 2000). To understand current structure we employ electro-technical model of current generator and simplify (1) by considering only convection term which we divide into two parts, $c\vec{E} = \vec{e}_y \cdot V_x B_z - \vec{e}_x \cdot V_y B_z$. The first part $E_y = V_x B_z/c$ drives current in Y direction at the equator and oppositely at the high-latitude magnetosheath where $B_z$ is negative. This is Chapman-Ferraro current and corresponding dynamic equation is given by

$$\frac{\partial J_y}{\partial t} \approx \frac{c}{4\pi}\frac{\partial^2 V_x B_z}{\partial x^2} \qquad (6)$$

The asymmetric part $E_x = -V_y B_z / c$ constitutes two opposite generators at the dawn and dusk sides of the equator. It was considered in relation to FAC as early as in (Eastman 1976). The dawn side generator drives current across the magnetosheath into the inner magnetosphere, and dynamic equation is given by

$$\frac{\partial J_x}{\partial t} \approx -\frac{c}{4\pi}\frac{\partial^2 V_y B_z}{\partial z^2} \quad (7)$$

In the inner magnetosphere this current goes along field lines to the dipole pole, while in the magnetosheath from high latitudes down to the equator, as given by (5). The ensuing picture is depicted in figure 11. We note that the constructed FAC loop is qualitatively very similar to results of a number of numerical simulations (Janhunen and Koskinen 1997, Siscoe *et al* 2002b). In terms of current generator it is clear that inside it FAC decelerates plasma flanking motion due to magnetic force $\sim J_x B_z$. Simple estimation shows that this force becomes significant when the total FAC value on both poles becomes comparable to the total Chapman-Ferraro current.

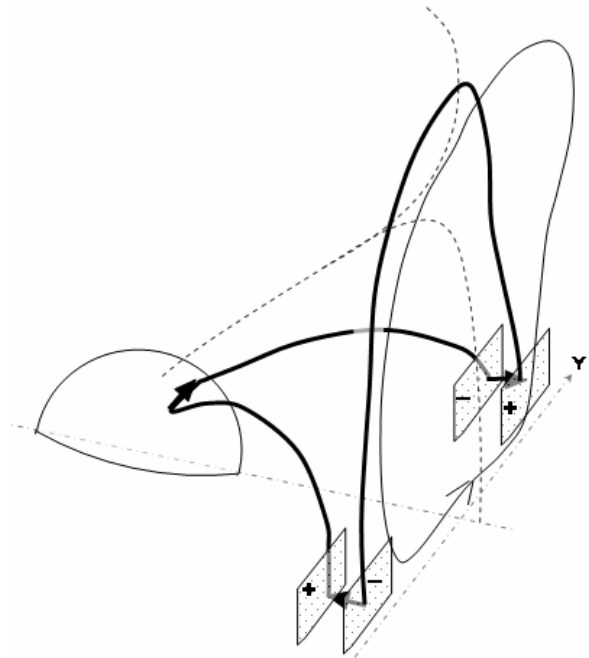

**Figure 11.** A sketch of FAC loop in the northern hemisphere. FAC generators are schematically shown as plates with respective charge sings. A thin solid line represents Chapman-Ferraro current, dashed - null magnetic field line.

### 4. Conclusions
In comparison to terrella, the reported experiment with dipole of large moment and laser-produced plasma enabled to increase the scale of laboratory magnetosphere, to observe intense field aligned current and to study it in detail for the first time. Obtained body of data shows analogy of observed current to Region-1 current on the Earth in such features as direction of flow, relative value and details of electron motion. A direct comparison of conductive and non-conductive dipole surfaces has been made for the first time. Thus, the specific input of FAC into magnetospheric field and the link between FAC and electric potential in equatorial boundary layer were revealed. The model of electric potential generation in the boundary layer which was considered earlier for calculating transpolar potential value in the Earth magnetosphere in absence of IFM qualitatively and quantitatively explains results of reported measurements. However, a feedback influence of FAC on electric potential was observed in the experiment which is a step forward. Due to non-stationary conditions of experiment results of the paper could be most relevant to initial magnetospheric response on sudden variations in SW pressure. The model presented in figure 11 gives indication how in future studies a mechanism of FAC generation and saturation could be elucidated. Another interesting case for future study is experiment with external magnetic field representing southward and northward IFM.

Acknowledgements: This work was supported by SB RAS Research Program, Russian Fund for Basic Research, grant 09-02-00492 and OFN RAS Research Program 16.

**References**

Antonov V M, Zakharov Yu P, Melekhov A V, Oraevskii V N, Ponomarenko A G and Posukh V G 2001 Explosion with plasma quasitrapping in a dipole field *Journal of Applied Mechanics and Technical Physics* **42** N6 949-58

Anderson B J, Acuna M H, Korth H, Purucker M E, Johnson C L, Slavin J A, Solomon S C, McNutt R L Jr 2008 The structure of Mercury's magnetic field from Messenger's first flyby *Science* **321** 82-5

Baumjohann W, Matsuoka A, Glassmeier K H, Russell C T, Nagai T, Hoshino M, Nakagawa T, Balogh A, Slavin J A, Nakamura R and Magnes W 2006 The magnetosphere of Mercury and its solar wind environment: Open issues and scientific questions *Advances in Space Research* **38** N4 604-9

Boyle, C B, Reiff P H and Hairston M R 1997 Empirical polar cap potentials *J. Geophys. Res.* **102**(A1) 111-25

Denisenko V V, Erkaev N V, Kitaev A V and Matveenkov I T 1992 *Mathematical Modeling of Magnetospheric Processes* [in Russian], ed V G Pivovarov (Novosibirsk: Nauka) pp. 62-78

Eastman, T E 1976 The magnetospheric boundary layer: site of plasma, momentum and energy transfer from the magnetosheath into the magnetosphere *Geophys. Res. Lett.* **3**(11) 685-8

Fedder, J A and Lyon J G 1987 The solar wind–magnetosphere–ionosphere current–voltage relationship *Geophys. Res. Lett.* **14** 880-3

Hasegawa A and Sato T 1979 Generation of field aligned current during substorm *Dynamics of the Magnetosphere,* ed S I Akasofu (Mass.: D. Reidel, Hingham) pp 529–42

Hill T W 1984 Magnetic coupling between solar wind and magnetosphere: Regulated by ionospheric conductance *Eos Trans. AGU* **65** 1047-8

Iijima T and Potemra T A 1976 The amplitude distribution of field-aligned currents at northern high latitudes observed by TRIAD *J. Geophys. Res.* **81** 2165-74

Itonaga M, Yoshikawa A and Fujita S 2000 A wave equation describing the generation of field-aligned current in the magnetosphere *Earth Planets Space* **52** 503–7

Janhunen P and Koskinen H E J 1997 The Closure of Region-1 Field-Aligned Current in MHD Simulation, *Geophys. Res. Lett.* **24**(11) 1419–22

Marchaudon A, Cerisier J C, Bosqued J M, Owen C J, Fazakerley A N and Lahiff A D 2006 On the structure of field-aligned currents in the mid-altitude cusp *Ann. Geophys.* **24** 3391–401

Nikitin S A and Ponomarenko A G 1995 Energetic criteria of artificial magnetosphere formation *Journal of Applied Mechanics and Technical Physics* **36** N4 483-7

Paschmann G, Schwartz S J, Escoubet C P, Haaland S 2005 Outer Magnetospheric Boundaries: Cluster Results *Space Sci. Rev.* **118** 231–424

Ponomarenko, A G, Antonov V M, Posukh V G, Melekhov A V, Boyarintsev E L, Afanasyev D M, Yurkov R N 2002 Simulation of non-stationary processes in the solar wind and its impact on the Earth magnetosphere *Report on MinPromNauka project "Investigation of Solar activity and its manifictations in near Earth space and atmosphere"* [in Russian] part III

Ponomarenko A G, Zakharov Yu P, Antonov V M, Boyarintsev E L, Melekhov A V, Posukh V G, Shaikhislamov I F and Vchivkov K V 2007 Laser Plasma Experiments to Simulate Coronal Mass Ejections During Giant Solar Flare and Their Strong Impact onto Magnetospheres *IEEE Trans. Plasma Sci.* **35** Issue 4 Pt. 1 813-21

Ponomarenko A G, Zakharov Yu P, Antonov V M, Boyarintsev E L, Melekhov A V, Posukh V G, Shaikhislamov I F and Vchivkov K V 2008 Simulation of strong magnetospheric disturbances in laser-produced plasma experiments *Plasma Phys. Contr. Fusion* **50** #074015 (10 pp)

Potter A E and Morgan T H 1990 Evidence for magnetospheric events on the sodium atmosphere of Mercury *Science* **248** 835

Rahman H U, Yur G, White R S, Birn J and Wessel F J 1991 On the Influence of the Magnetization of a Model Solar Wind on a Laboratory Magnetosphere *J. Geophys. Res.*, **96**(A5) 7823–9

Ridley A J, De Zeeuw D L, Manchester W B and Hansen K C 2006 The magnetospheric and ionospheric response to a very strong interplanetary shock and coronal mass ejection *Advances in Space Research* **38** N2 263-72

Shepherd Simon G 2007 Polar cap potential saturation: Observations, theory, and modeling *J. of Atmospheric and Solar-Terrestrial Phys.* **69** N3 234-48

Siscoe G L, Erickson G M, Sonnerup B U O, Maynard N C, Schoendorf J A, Siebert K D, Weimer D R, White W W and Wilson G R 2002a Hill model of transpolar potential saturation: Comparisons with MHD simulations *J. Geophys. Res* **107** A6 1075

Siscoe G L, Crooker N U and Siebert K D 2002b Transpolar potential saturation: Roles of region 1 current system and solar wind ram pressure *J. Geophys. Res* **107** A10 1321

Slavin J A, Owen J C J, Connerney J E P and Christon S P 1997 Mariner 10 observations of field-aligned currents at Mercury *Planet. Space Sci.* **45** 133–41

Vasyliunas V M 1984 Fundamentals of current description *Magnetospheric Currents*, ed T A Potemra (Washington D.C.: AGU) pp 63–6

Zakharov Yu P, Antonov V M, Boyarintsev E L, Melekhov A V, Posukh V G, Shaikhislamov I F, Vchivkov K V, Nakashima H and Ponomarenko A G 2008 New Type of Laser-Plasma Experiments to simulate an Extreme and Global Impact of Giant Coronal Mass Ejections onto Earth' Magnetosphere *J. Phys.: Conf. Ser.* **112** #042011